\documentclass[aps,prl,preprint,superscriptaddress,tightenlines]{revtex4}

\usepackage{graphicx}
\usepackage{dcolumn}
\usepackage{bm}

\begin{document}

\preprint{\vbox{\hbox{\hfil CLNS 01/1767}
                        \hbox{\hfil CLEO 01-23}
}}

\title{Improved Measurement of $|V_{ub}|$ with Inclusive Semileptonic $B$ Decays}

\author{A.~Bornheim}
\author{E.~Lipeles}
\author{S.~P.~Pappas}
\author{A.~Shapiro}
\author{W.~M.~Sun}
\author{A.~J.~Weinstein}
\affiliation{California Institute of Technology, Pasadena, California 91125}                
\author{G.~Masek}
\author{H.~P.~Paar}
\affiliation{University of California, San Diego, La Jolla, California 92093}               
\author{R.~Mahapatra}
\author{R.~J.~Morrison}
\affiliation{University of California, Santa Barbara, California 93106}                     
% ...
\collaboration{CLEO Collaboration}
\noaffiliation

\date{February 7, 2002}

\begin{abstract}

We report a new measurement of the CKM parameter $|V_{ub}|$
made with a sample of 9.7 million $B {\bar B}$ events collected 
with the CLEO~II detector.  Using Heavy Quark theory, we combine the 
observed yield of leptons from semileptonic $B$ decay in the end-point 
momentum interval $2.2-2.6$~GeV/$c$ with recent CLEO~II  
data on $B \rightarrow X_s \gamma$ to find 
$|V_{ub}| = (4.08 \pm 0.34 \pm 0.44 \pm 0.16 \pm 0.24) \times 10^{-3}$, 
where the first two uncertainties are experimental and the last two are 
from theory.

\end{abstract}



\maketitle

Measurement of the Cabibbo-Kobayashi-Maskawa matrix element $V_{ub}$ is 
one of the most challenging and important tasks in testing the Standard 
Model.  This direct measurement of one side of the unitarity triangle
gives information that is complementary to studies of CP violation in 
$B$ decays. Semileptonic $B$ decays are a direct route to 
$|V_{ub}|$ and $|V_{cb}|$.  Measurements of lepton production above the 
kinematic limit for $B \rightarrow X_c \ell \nu$ 
provided the discovery of nonzero 
$|V_{ub}|$ \cite{Fulton:1990pk,Albrecht:1990zf}
and a measurement of its value \cite{Bartelt:1993xh} that was controversial
due to its heavy reliance on theoretical models.

In this paper we report an improved inclusive measurement of 
$B \rightarrow X_u \ell \nu$ made with the CLEO~II detector 
at the Cornell Electron Storage Ring (CESR).  This measurement
supersedes that of Ref.~\cite{Bartelt:1993xh}, which was based on the
first tenth of our data sample.   
Improved background-suppression techniques help to reduce the model
dependence of the result.  We have also measured 
$B \rightarrow X_u \ell \nu$ over a broader momentum range than was previously 
possible, further helping to reduce theoretical uncertainty, but incurring
experimental uncertainty because of the large $B \rightarrow X_c \ell
\nu$ subtraction.  We use the momentum range $2.2-2.6$~GeV/$c$ as a  
compromise between these considerations, and present other intervals to 
probe the stability of our result and to compare with past measurements. 
For extracting $|V_{ub}|$, models have been replaced with Heavy Quark (HQ) 
theory 
\cite{Neubert:1994ch,Neubert:1994um,Bigi:1994ex,Bigi:1994it,Kagan:1998ym,
DeFazio:1999sv,Leibovich:1999xf} and the CLEO-measured
$B \rightarrow X_s \gamma$ photon-energy spectrum \cite{Chen:2001fj}. 
These improvements result in a determination
of $|V_{ub}|$ with overall uncertainty that is smaller and more 
reliably estimated than previous measurements.

The CLEO~II detector has been described in detail \cite{Kubota:1992ww}.
Key components for this work are the tracking system, the cesium iodide 
calorimeter, and the muon detector.  Two thirds of our data were
collected after a detector upgrade that included a silicon 
tracker \cite{Hill:1998ea}, a change of drift-chamber gas from argon-ethane 
to helium-propane, and other improvements.
We obtained an integrated luminosity of 9.13~fb$^{-1}$ of $e^+e^-$ 
annihilation data at the $\Upsilon(4S)$ resonance (ON) and 4.35~fb$^{-1}$ 
at energies just below $B {\bar B}$ threshold (OFF).  OFF yields are scaled 
by a factor of $2.06 \pm 0.02$ and subtracted from the ON yields to
eliminate contributions of non-$B\bar{B}$ processes.
We determine the scale factor by comparing ON and OFF track spectra 
above the $B$-decay kinematic limit, and confirm the result with measured 
luminosities.  The ON sample includes 9.7~million $B {\bar B}$ events.  

We select electron and muon candidates in the central region of CLEO~II 
($|\cos \theta| \leq 0.71$, where $\theta$ is the angle with respect to 
the beam).  An electron candidate must have an energy
deposit in the calorimeter close in value to its measured momentum and a
specific
ionization consistent with that expected for an electron.  The efficiency of 
electron selection is slightly momentum-dependent, with a value of 
$0.93 \pm 0.03$ at 2.2~GeV/$c$.  Muons must penetrate seven 
nuclear interaction lengths of absorber material.  The threshold momentum 
is $\sim 1.8$~GeV/$c$ and the efficiency is $0.88 \pm 0.03$ at 2.2~GeV/$c$.  
Misidentification rates are measured for tagged $\pi^\pm$, $K^\pm$, and 
$p/\bar{p}$ tracks 
and averaged in proportions determined with simulated $B {\bar B}$ events.  
The electron and muon fake rates above 2.0 GeV/$c$ are less than 0.1\% and 
about 0.5\%, respectively.

We use several cuts to minimize mismeasurements and background from
sources other than semileptonic decays.  We veto any lepton that can be 
paired with another lepton of the same type and opposite charge if the pair 
mass is within three standard deviations of the $J/\psi$ mass.  Electrons 
from photon conversions are also rejected. Stringent requirements on tracking 
residuals, impact parameter, and the fraction of tracking layers traversed 
that have good hits ensure a reliable momentum measurement.  The 
momentum-dependent track-selection efficiency is determined with a GEANT 
simulation \cite{GEANT} of signal events.  Real and simulated radiative 
Bhabha electrons embedded in hadronic events are used to determine a
correction factor
accounting for the imperfect simulation.  The track-selection efficiency 
for $B \rightarrow X_u \ell \nu$ is $\sim 0.95$, with a 3\% uncertainty.

Although most leptons at the $\Upsilon(4S)$ are from semileptonic 
$B$ decays, continuum production contributes a large background at high
momentum that can be subtracted with OFF data.  To avoid degrading the
precision of our $B \rightarrow X_u \ell \nu$ measurement, we use
several criteria  to suppress the continuum before subtraction.
Beam-gas, beam-wall, and some QED events are  
eliminated by cuts on visible energy and the event vertex.  A multiplicity 
cut suppresses $\tau^+ \tau^-$.  Particles above the $B$-decay kinematic 
limit indicate continuum production or mismeasurement, so we eliminate 
events having a track or shower with an energy above 3.5~GeV.
Two-photon and QED processes are suppressed by cutting events with
missing momentum near the beam. 
\begin{table*} [tbhp]
\caption[tab:yields2]{Lepton yields and backgrounds in the momentum interval 
$2.2-2.6$~GeV/$c$.}
\label{tab:yields2}
\begin{ruledtabular}
\begin{tabular}{lccc}
& $e$ & $\mu$ & Sum \\
\hline
$N_{ON}$ & 4110 & 4857 & 8967 \\
$N_{OFF}$ & 410 & 573 & 983 \\
\hline
Excess                       & $3265 \pm 77 \pm 8$   & $3673 \pm 85 \pm 12 $ & $6938 \pm 115 \pm 20 $ \\
Fakes                        & $15   \pm 6  \pm 4$   & $194 \pm 13 \pm 58 $ & $209  \pm 19  \pm 58 $ \\
$J/\psi$                     & $68   \pm 4  \pm 7$   & $90  \pm 5  \pm 9  $ & $158  \pm 6   \pm 16 $ \\
Other Backgrounds            & $40   \pm 8  \pm 10$  & $67  \pm 6  \pm 18 $ & $107  \pm 10  \pm 29 $ \\
$B \rightarrow X_c \ell \nu$   & $2147 \pm 23 \pm 116$ & $2415 \pm 24 \pm 130$ & $4562 \pm 33  \pm 246$ \\
\hline
$B \rightarrow X_u \ell \nu$ & $995  \pm 81 \pm 117$ & $906 \pm 106 \pm 133$ & $1901 \pm 122 \pm 256$ \\
\end{tabular}
\end{ruledtabular}
\end{table*}

The continuum background remaining after this selection is
primarily $e^+e^- \rightarrow c {\bar c}$.  In our previous 
analysis \cite{Bartelt:1993xh}, cuts on 
event-shape variables and missing momentum used the 
kinematics of $B \rightarrow X_u \ell \nu$ to suppress the continuum 
by a factor of 70, with a signal efficiency of $\sim 0.38$. 
The efficiency depended strongly on $q^2$ (the squared mass of the
virtual $W$), and thus
contributed to the model dependence of the result.  The goal of this
analysis is to achieve comparable suppression
with reduced model dependence.  This requires cuts that exploit the 
presence of the ``other $B$'' in signal events rather than the details 
of $B \rightarrow X_u \ell \nu$.  We use a neural 
net with inputs that are measurements of the energy flowing into eleven
angular intervals defined with respect to the candidate-lepton direction.  
We exclude energy within 25$^\circ$ of the direction 
opposite to the lepton, where there is strong dependence on the $q^2$ of 
the decay.  The net was trained with a 
simulated signal sample generated with the ISGW2 model \cite{Scora:1995ty} 
and a simulated background sample of continuum 
$e^+e^- \rightarrow $~$q {\bar q}$. Optimized 
for expected signal level, the neural net  
gives background rejection of a factor of 50 and signal efficiency of 0.33.

The total efficiency for selecting leptons from $B \rightarrow X_u \ell \nu$
($\sim 0.21$) shows a small momentum dependence over the range 
$2.0-2.6$~GeV/$c$.  The uncertainty is about 7\%, with roughly equal
contributions from detector response and model uncertainties.  We assess 
the latter by considering several models: ISGW2, the ACCMM spectator model 
\cite{Altarelli:1982kh,Artuso:1993fb} with various values of internal 
parameters, and a hybrid of these developed by CLEO.  Our current 
procedures have less than one third of the model-to-model variation of 
the previous analysis and essentially no 
dependence on the mass of the hadronic system $X_u$.

The computation of the $B \rightarrow X_u \ell \nu$ signal for the momentum
interval $2.2-2.6$~GeV/$c$ is shown in Table~\ref{tab:yields2}.
The spectrum of lepton candidates from $B {\bar B}$ 
events is obtained by subtracting the scaled OFF momentum distribution from 
the ON distribution.  It includes $B \rightarrow X_u \ell \nu$ and several 
$B$-decay backgrounds.  Hadrons misidentified as leptons 
(fakes) are computed by combining the momentum-dependent misidentification
probabilities with the spectrum of hadrons from $B {\bar B}$ that pass all
selection requirements except for lepton identification.  Leakage through the 
$J/\psi$ and photon-conversion vetoes is estimated with Monte Carlo 
normalized to data, with systematic errors of 10\% and 25\%, respectively.  
Several other backgrounds are also estimated by Monte 
Carlo and are assigned conservative systematic errors: $\psi(2S)$ 
leptonic decays ($\pm 25\%$), semileptonic and leptonic 
decays of $D$ and $D_s$ ($\pm 50\%$), semileptonic decays 
to $\tau$ ($\pm 25\%$), and $\pi^0$ Dalitz decays ($\pm 100\%$).

The largest background to $B \rightarrow X_u \ell \nu$ for
momenta below 2.4~GeV/$c$ is $B \rightarrow X_c \ell \nu$.  We
calculate this by fitting the lepton spectra between 1.5~GeV/$c$ and the 
low end of the end-point interval.  Fit functions are generated with Monte
Carlo simulations using models and CLEO-measured form 
factors \cite{Duboscq:1996mv,Athanas:1997eg}.  QED radiative corrections are 
applied with the PHOTOS algorithm \cite{Barberio:1994qi}.  Spectra generated 
in the $B$ rest frame are boosted to the lab frame using the $B$-momentum 
distribution of our data.  There are four components in the fits: a mixture 
of $D$ and $D^*$ with the ratio given by exclusive branching 
fractions \cite{Groom:2000in}; a mixture of decays to $D^{**}$ and other 
higher-mass charmed mesons (ISGW2); nonresonant decays (model of Goity and 
Roberts \cite{Goity:1995xn}); and $B \rightarrow X_u \ell \nu$ (ISGW2 
normalized to the end-point yield of Ref.~\cite{Bartelt:1993xh}).  

The proportions for the $B \rightarrow X_c \ell \nu$ components are
determined by fits to the electron spectrum without the neural-net cut,
because the momentum acceptance is larger and has less uncertainty than
that for the muons.  The resulting fit has a $\chi^2$ of 
14.9 for 11 degrees of freedom.  
The muons are then fitted to the same mixture, 
with one parameter to allow for a difference in the $e/\mu$ efficiency 
ratio between Monte Carlo and data.  The electron and muon spectra with the 
neural-net cut applied are then fitted to the same mixtures, with one 
parameter allowing for mis-modeling of the neural-net efficiency.  
These fits provide the $B \rightarrow X_c \ell \nu$ subtractions in 
Table~\ref{tab:yields2}.

We assess the uncertainty in the $B \rightarrow X_c \ell \nu$ subtraction 
by varying inputs, including the $D/D^*$ 
ratio, the form factors for $B \rightarrow D/D^* \ell \nu$, the relative
$D^{**}$ and nonresonant normalizations, the radiative corrections, the
$B$-momentum scale, the normalization and models for 
$B \rightarrow X_u \ell \nu$, and the fit intervals.  The overall
uncertainty in the subtraction is the sum in quadrature of the observed
variations.  The largest is due to form factors.

\begin{figure}[hbtp]
\begin{center}
\resizebox{3 in}{!}{
\includegraphics{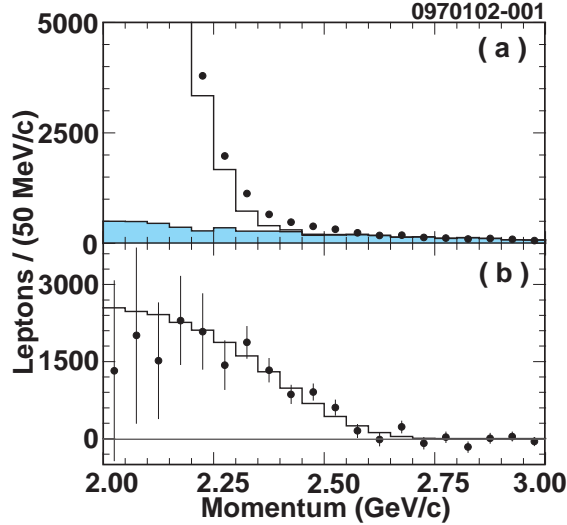}
}
\caption{(a) Lepton spectra for ON (points) and scaled OFF
(shaded histogram).  The unshaded
histogram is the sum of the scaled OFF and $B$-decay backgrounds.  
(b) Background-subtracted efficiency-corrected lepton spectrum for 
$B \rightarrow X_u \ell \nu$ (points).  The histogram is the 
$B \rightarrow X_u \ell \nu$ spectrum predicted with the measured 
$B \rightarrow X_s \gamma$ spectrum.}
\label{fig:endpoint}
\end{center}
\end{figure}

Fig.~\ref{fig:endpoint}(a) shows the ON and scaled OFF momentum spectra.
Fig.~\ref{fig:endpoint}(b) shows the 
background-subtracted and efficiency-corrected spectrum for $B
\rightarrow X_u \ell \nu$. 
Statistical and systematic uncertainties have been combined in 
quadrature.  Below 2.3~GeV/$c$, the 
$B \rightarrow X_c \ell \nu$ subtraction dominates the uncertainties,
which are strongly correlated from bin to bin.
The partial branching fraction for $B \rightarrow X_u \ell \nu$
($\ell$ = $e$ or $\mu$) in a given momentum interval 
is $\Delta {\cal B}_u(p) = N_\ell(p)/[2 N_{B {\bar B}} \epsilon(p)]$,
where $N_\ell$ is half of the total yield of $e$'s and $\mu$'s, 
$\epsilon(p)$ is the efficiency, and 
$N_{B {\bar B}}=(9.67\pm 0.17) \times 10^6$ 
is the number of $B {\bar B}$ events.
Results are given in Table~\ref{tab:results} for five different
end-point momentum intervals.
\begingroup
\squeezetable
\begin{table*} [tbh]
\caption[tab:results]{Results for five momentum intervals.  Uncertainties 
on yields, $f_u$, and branching fractions are statistical and 
systematic.  The first uncertainty on the total 
branching fraction is from the measurement of $\Delta {\cal B}_u(p)$ and
the second is from $f_u$.  The first two uncertainties on $|V_{ub}|$ are
from the branching fraction and the third and fourth are from theory.}
\label{tab:results}
\begin{ruledtabular}
\begin{tabular}{cccccc}
$p$ (GeV/$c$) &	Yield	&	$\Delta {\cal B}_u(p)$($10^{-4}$) &	$f_u$	& ${\cal B}(B\rightarrow X_u \ell \nu)$ ($10^{-3}$) & $|V_{ub}|$($10^{-3}$) \\
\hline
2.0-2.6	& $3538 \pm 279 \pm 1470$  & $4.22 \pm 0.33 \pm 1.78$   & $0.266 \pm 0.041 \pm 0.024$   & $1.59 \pm 0.68 \pm 0.28$	  & $3.87 \pm 0.83 \pm 0.35 \pm 0.15 \pm 0.12$ \\
2.1-2.6	& $2751 \pm 191 \pm  584$  & $3.28 \pm 0.23 \pm 0.73$   & $0.198 \pm 0.035 \pm 0.020$   & $1.66 \pm 0.39 \pm 0.34$	  & $3.95 \pm 0.46 \pm 0.40 \pm 0.16 \pm 0.16$ \\
2.2-2.6	& $1901 \pm 122 \pm  256$  & $2.30 \pm 0.15 \pm 0.35$   & $0.130 \pm 0.024 \pm 0.015$   & $1.77 \pm 0.29 \pm 0.38$	  & $4.08 \pm 0.34 \pm 0.44 \pm 0.16 \pm 0.24$ \\
2.3-2.6	& $1152 \pm  80 \pm   61$  & $1.43 \pm 0.10 \pm 0.13$   & $0.074 \pm 0.014 \pm 0.009$   & $1.94 \pm 0.22 \pm 0.43$	  & $4.27 \pm 0.24 \pm 0.47 \pm 0.17 \pm 0.34$ \\
2.4-2.6	& $499  \pm  57 \pm   14$  & $0.64 \pm 0.07 \pm 0.05$   & $0.037 \pm 0.007 \pm 0.003$   & $1.74 \pm 0.24 \pm 0.38$	  & $4.05 \pm 0.28 \pm 0.45 \pm 0.16 \pm 0.45$ \\
\end{tabular}
\end{ruledtabular}
\end{table*}
\endgroup
They are in good agreement with our previous
measurement~\cite{Bartelt:1993xh}.

To determine the charmless semileptonic branching fraction 
${\cal B}(B \rightarrow X_u \ell \nu)$ from the partial 
branching fraction $\Delta {\cal B}_u(p)$, we need to know the true fraction 
$f_u(p)$ of the $B \rightarrow X_u \ell \nu$ spectrum that falls in the 
given momentum interval.
Parton-level decays can be reliably calculated, but the observable
meson-decay processes depend on the mass of the $b$ quark and its motion
inside the $B$ meson.  These properties have been described in recent
years with HQ theory
\cite{Neubert:1994ch,Neubert:1994um,Bigi:1994ex,Bigi:1994it,Kagan:1998ym,
DeFazio:1999sv,Leibovich:1999xf}.  A light-cone shape  
function can be convoluted with the parton-level $b \rightarrow u \ell \nu$
to obtain the spectrum for $B \rightarrow X_u \ell \nu$.  The shape
function depends on nonperturbative QCD and has not yet been calculated
from first principles.  To leading order, the same shape function describes
{\em all} $b$-to-light transitions, in particular also relating $B
\rightarrow X_s \gamma$ to $b \rightarrow s \gamma$.

We have recently measured the photon-energy spectrum in
$B \rightarrow X_s \gamma$ \cite{Chen:2001fj}. Three two-parameter 
descriptions of the shape function \cite{Bigi:1994it,Kagan:1998ym} are
used to fit this spectrum over the range $1.5 < E_\gamma < 2.8$~GeV.  
All shape functions give good fits.  We use the best-fit parameters and error 
ellipses to calculate the $B \rightarrow X_u \ell \nu$ 
lepton-momentum spectra and determine $f_u$ and its uncertainty, following
Ref.~\cite{DeFazio:1999sv}.  We compute the effect of QED radiative
corrections on $f_u$ with PHOTOS.  The resulting values of $f_u$ are given 
in Table~\ref{tab:results}.  
The systematic uncertainty includes contributions from the subtraction
of $B$-decay processes other than $B \rightarrow X_s \gamma$, the choice of
scale for evaluating $\alpha_s$, radiative corrections, and differences
among the shape functions.

We extract $|V_{ub}|$ by averaging the nearly identical
formulations of Hoang {\it et al.} \cite{Hoang:1998hm} and
Uraltsev \cite{Uraltsev:1999rr}:
\begin{displaymath}
|V_{ub}| = (3.07 \pm 0.12) \times 10^{-3} \times
\biggl[\frac{{\cal B}(B \rightarrow X_u e \nu)} {0.001}
\frac{1.6 {\rm~ps}} {\tau_B}\biggr]^{1 \over 2}.
\end{displaymath}
For the $B$ lifetime we use $\tau_B = 1.60 \pm 0.02$~ps \cite{Groom:2000in}.  
Results for $|V_{ub}|$ (Table~\ref{tab:results})
show excellent agreement among the five momentum ranges.  The best overall 
precision (15\%) is obtained for the $2.2-2.6$~GeV/$c$ interval.  The first 
uncertainty is from the measurement of $\Delta {\cal B}_u(p)$ (combined
statistical and systematic).  The second is the combined uncertainty in the
determination of $f_u$ from $B \rightarrow X_s \gamma$.  The third is
the average of the uncertainties given in
Refs.~\cite{Hoang:1998hm,Uraltsev:1999rr} for  
the extraction of $|V_{ub}|$ from the branching fraction.  The
fourth is the theoretical uncertainty associated with the assumption
that $B \rightarrow X_s \gamma$ can be used to compute the  
spectrum for  $B \rightarrow X_u \ell \nu$.  This is valid to leading
order, with corrections at order $\Lambda_{QCD}/M_B$.  Taking
$\Lambda_{QCD}/M_B \approx 0.1$, we estimate this error by varying the
shape function parameters by $\pm$10\% \cite{Neubert_PC}.  Uncertainty
due to the theoretical assumption of quark-hadron duality remains
unquantifiable. 

For comparison, we also determine $f_u$ with models of 
$B \rightarrow X_u \ell \nu$.  Available exclusive models are limited in 
the final states included, while inclusive models have uncertain 
internal parameters.  Neither type provides 
a solid basis for assessing the theoretical uncertainty.  For ISGW2 and ACCMM 
(spectator mass $m_{sp}$=150~MeV/$c^2$, Fermi momentum
$p_F$=300~MeV/$c$), we find  $|V_{ub}|$ in the $2.2 - 2.6$~GeV/$c$
interval to be 20\% smaller than the  result given above. Replacing the
default ACCMM parameters (chosen for consistency  with past analyses)
with values obtained by fitting the  $B \rightarrow X_s \gamma$ spectrum
to a spectator  parameterization \cite{Ali:1995bi} gives better
agreement.  The parameter  values are $m_{sp}\simeq$230~MeV/$c^2$ and $p_F
\simeq$440~MeV/$c$, leading to a  $|V_{ub}|$ that is $\sim$5\% smaller
than our result. 

In conclusion, we have measured the CKM parameter $|V_{ub}|$ to be
$(4.08 \pm 0.63) \times 10^{-3}$.  This result has smaller 
overall uncertainty than previous measurements and represents a major 
step forward, both in the quality of the experimental data and in the 
use of QCD theory rather than phenomenological models.


We gratefully acknowledge the efforts of the CESR staff in providing our
excellent data sample.  We thank A. Falk, A. Kagan, 
A. Leibovich, M. Luke, M. Neubert, I. Rothstein, M. Shifman, A. Vainshtein, 
and M. Voloshin for useful correspondences and discussions.  This work
was supported by the National Science Foundation, the U.S. Department
of Energy, the Research Corporation, and the Texas Advanced Research Program.


\end{document}